\documentclass[12pt,article]{revtex4}
\usepackage{epsfig}
\usepackage{bbm}

\begin{document}

\rightline{HD-THEP-04-35, ITP-UU-04/27, SPIN-UU-04/15}

\title{Coherent Baryogenesis and \\Nonthermal Leptogenesis: A comparison
\footnote{Contribution to the Proceedings of Strong and Electroweak Matter 2004
(SEWM2004), Helsinki, Finland, June 16-19, 2004.}}

\author{Bj\"orn Garbrecht, Tomislav Prokopec\\and Michael G. Schmidt}

\address{Institut f\"ur Theoretische Physik, Heidelberg University,
             Philosophenweg 16, D-69120 Heidelberg, Germany\\
E-mail:B.Garbrecht-,  T.Prokopec-, M.G.Schmidt@thphys.uni-heidelberg.de}

\maketitle

\centerline{\bf Abstract}

We present a new mechanism for baryogenesis: 
at preheating after inflation fermions acquire a varying mass
 by their coupling to a time dependent field. Their
CP-violating mass matrix can generate a charge asymmetry
 to be transformed into a lepton asymmetry through decay into standard model
particles and heavy Majorana neutrinos.
 In a concrete model of hybrid inflation we compare ``coherent baryogenesis'' 
with nonthermal leptogenesis by perturbative decay of the inflation
condensates.

\section{Introduction}
\label{introduction}
The baryon asymmetry of the universe $n_B/s\simeq 7-8\times 10^{-11}$ 
(BAU), which is now deduced from early nucleosynthesis data and from WMAP data
on microwave background fluctuations, is in surprisingly good agreement. 
Its theoretical explanation requires detailed knowledge on
B,~C/CP violation and nonequilibrium and this is an ideal testing ground
 for elementary particle physics models. Electroweak baryogenesis
 \cite{1} in a strong first order phase transition requires a rather
 low Higgs mass. This and other ingredients can be tested
 in present and near future experiments - a great advantage. 
But this also led to the conclusion that the Standard Model (SM)
 is ruled out for such a mechanism and that the MSSM is close to this
 borderline. Out of equilibrium decay of heavy Majorana neutrinos
 - CP-violating at one loop level through a matrix of Yukawa couplings 
- has become very attractive again because such Majorana neutrinos also
 play an important role in explaining the newly found neutrino mass pattern.
 If one starts with thermalised Majorana neutrinos one can arrive at
 a simple and very restrictive picture \cite{2} though in the SUSY
 case there are some problems with gravitino overproduction. 
We here present a new `coherent baryogenesis' mechanism,
which also contains lepton number violation by Majorana neutrinos
 and compare it with nonthermal leptogenesis by perturbative decay of the
 inflaton field and with nonthermal leptogenesis during preheating.

\vspace{-0.2cm}
\section{Coherent baryogenesis, formalism, and a concrete model}
\label{coherentbaryogenesis}
At the end of inflation, there is an oscillating condensate which couples to fermions and induces a varying mass matrix. This, in turn, yields fermion production, known as preheating. Moreover, the mass matrix induces CP-violating flavour oscillations, which is a tree level effect. Charge is produced and is frozen in when the scalar condensate settles to its minimum. Eventually, this charge gets converted to {\em (B-L)}. This mechanism we call ``coherent baryogenesis'' \cite{3}. We use the following system of equations for charge and currents of a fermionic system derived in the Schwinger-Keldish formalism for two point functions after Wigner transform
\begin{eqnarray}\label{1}
&&\dot{f}_{0h} + i\left[M_H,f_{1h}\right] +i\left[M_A,f_{2h}\right] = 0
\nonumber\\
&&\dot{f}_{1h} + 2h|\mathbf{k}|f_{2h} + i\left[M_H,f_{0h}\right] -\left\{M_A,f_{3h}\right\}  = 0
\nonumber\\
&&\dot{f}_{2h} - 2h|\mathbf{k}|f_{1h}  + \left\{M_H,f_{3h}\right\} +i\left[M_A,f_{0h}\right] = 0
\nonumber\\
&&\dot{f}_{3h} - \left\{M_H,f_{2h}\right\} +\left\{M_A,f_{1h}\right\}  = 0
\end{eqnarray}
with $f_{0h}$: charge density, $f_{3h}$:  axial charge density, $f_{1h}$:  
scalar density, $f_{2h}$:  pseudoscalar density, $h$:  helicity, $k$  
momentum, and hermitean and antihermitean part of the mass matrix,
 $M_H=\frac{1}{2}(M+M^+)$ and $M_A=\frac{1}{2i}(M-M^+)$. 
For pure quantum states this is equivalent to using the Dirac equation for
 mixing fermions, but it is also a generalization for mixed states.

In order to demonstrate our  new proposal we  implement it in a realistic model for hybrid inflation \cite{4}, the supersymmetric Pati-Salam model with gauge group 
$G_{PS}=SU(4)_c\times SU(2)_L\times SU(2)_R$ which after modifications does not suffer from the monopole problem as discussed in \cite{5}. The relevant terms of the superpotential are 
\begin{equation}\label{3}
W=\kappa S(\bar{H}^cH^c-\mu^2)-\beta S\left(\frac{\bar{H}^cH^c}{M_s}\right)^2+
\zeta GH^cH^c+\xi G\bar{H}^c\bar{H}^c
\end{equation}
with superfields $H^c=(\bar{4},1,2),~\bar{H}^c=(4,1,2),~S=(1,1,1),~G=(6,1,1)$ and a notation 
$H^c=\left(\begin{array}{llll}
u^c_{H_1}&u^c_{H_2}&u^c_{H_3}&\nu^c_H\\
d^c_{H_1}&d^c_{H_2}&d^c_{H_3}&e^c_H
\end{array}\right)$. 
This leads to a scalar potential
\begin{equation}\label{5}
V=2\left|S\nu_H^{c*}\left(\kappa-
2\beta \frac{|\nu_H^c|^2}{M_S^2}\right)\right|^2
+\left|\kappa\left(|\nu_H^c|^2-\mu^2\right)-\beta\frac{|\nu_H^c|^4}{M_S^2}
\right|^2.
\end{equation}
During inflation SUSY is broken, the $s$neutrino like scalar Higgses have an expectation value 
$\langle|\nu_H^c|\rangle =M_S\sqrt{\kappa/2\beta}$. This field and the inflaton 
field $S$ begin to fall rapidly during the waterfall regime at the end
of inflation (see figures below) before the SUSY vacuum is approached at $S=0$. 
%
\begin{figure}[htbp]
\begin{center}
\epsfig{file=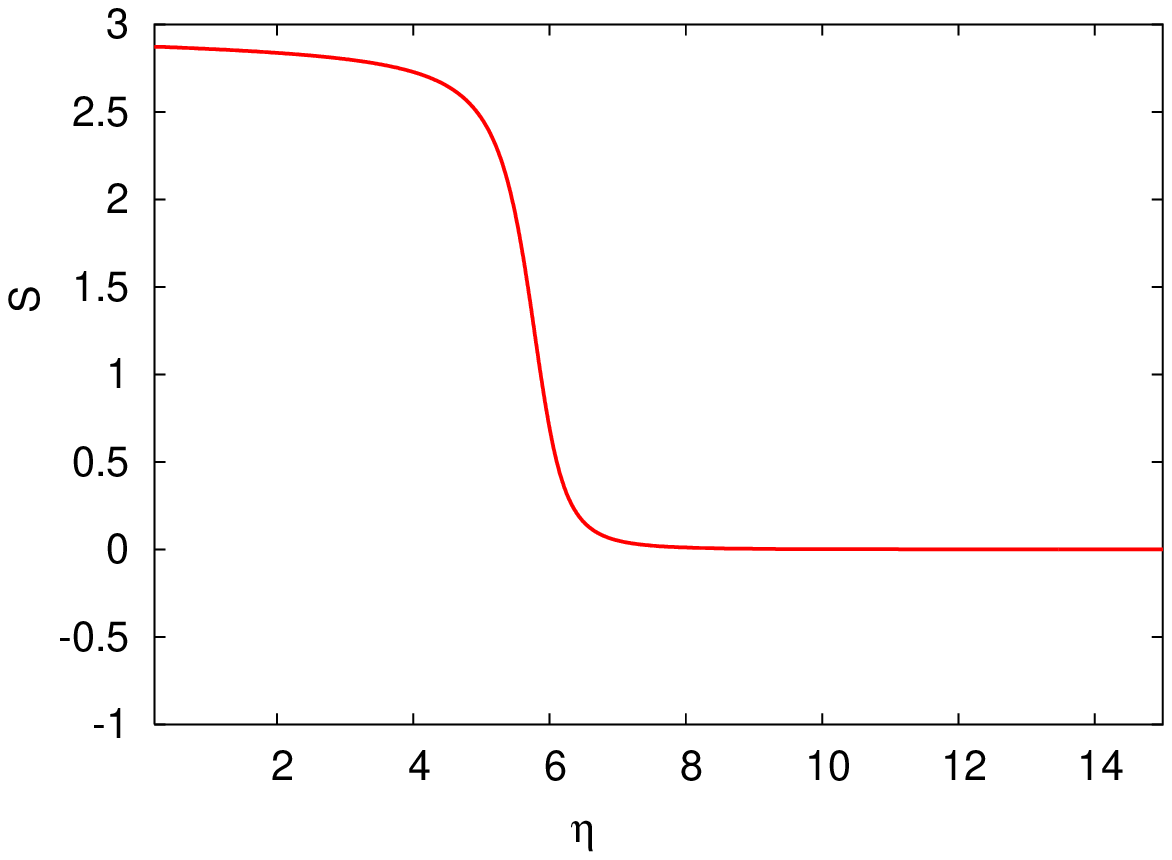, height=2.4in,width=3.2in}
\epsfig{file=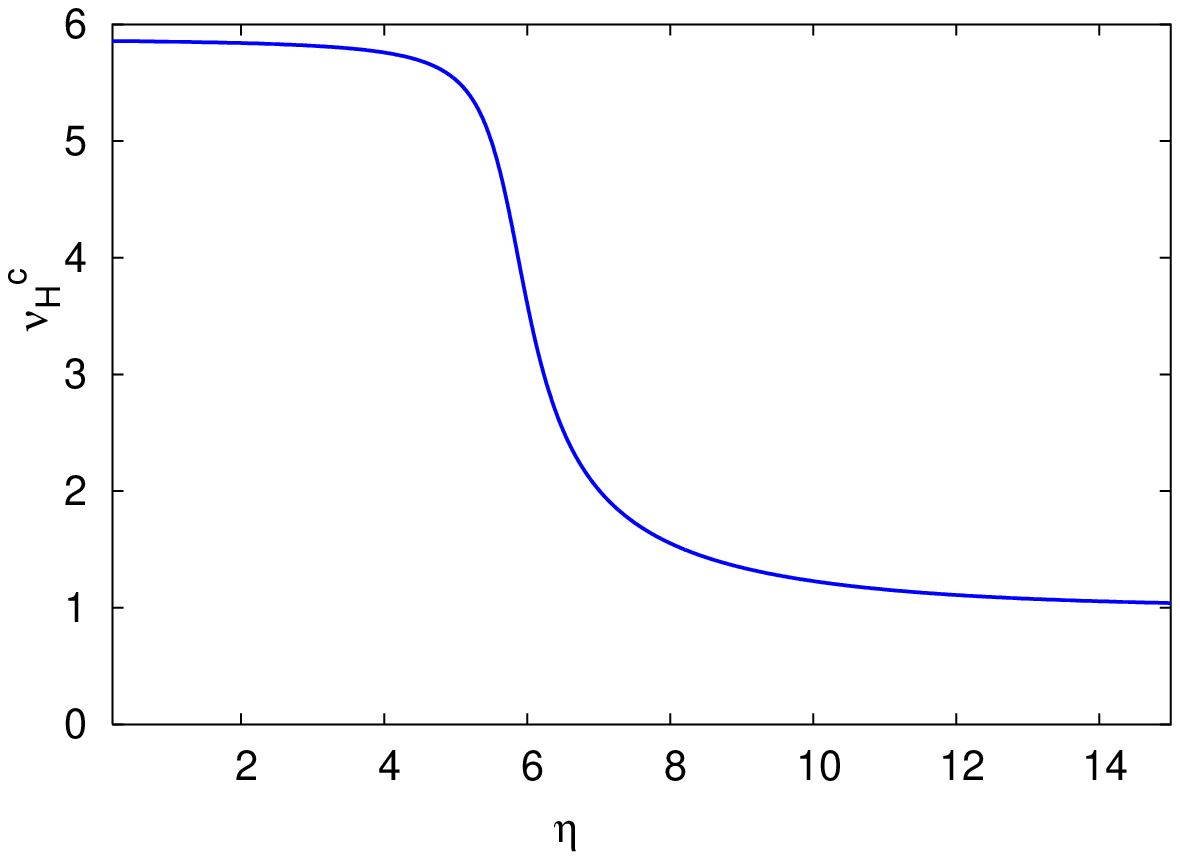, height=2.4in,width=3.2in}
\end{center}
\vskip -0.30in
\end{figure}
This induces a time dependent mass matrix for the Dirac fermions \qquad $\chi_{1j}=\left(-\Psi_{d^cH_j},~\bar{\Psi}_{\bar{D}_j}\right),
~\chi_{2_j}=\left(\Psi_{D_j},~\bar{\Psi}_{d^cH_j}\right)$
\begin{equation}
(\bar{\chi}_{1j},\bar{\chi}_{2j})\!
\left[\left(\begin{array}{cc}
\Re\left[\langle\nu^c_H\rangle\xi\right]&\frac{1}{2}m_d\\
\frac{1}{2}m_d&\Re\left[\langle\nu^c_H\rangle\zeta\right]\end{array}\right)
 \!+\!i\gamma_5\right.
\left.\!\!\left(\begin{array}{cc}
\!-\Im\left[\langle\nu^c_H\rangle\xi\right]&-\frac{i}{2}m_d
\\
\frac{i}{2}m_d&-\!\Im\left[\langle\nu^c_H\rangle\zeta\right]
 \end{array}\right)\right]
\left(\begin{array}{c}\chi_{1j}
\\ \chi_{2j}\end{array}\right)
,\nonumber
\end{equation}
where $m_d=\langle S\rangle
            \left(\kappa/2-\beta\langle|\nu^c_H|^2\rangle /M^2_S\right)$ 
and we have chosen the parameters $\kappa=0.007$, $\beta=1$, $\xi=0.12$,
$\zeta=0.12i$, $M_s=100\mu$, $\mu=3.9\times 10^{16}$~GeV,
 $\gamma=\gamma_1=0.0001$,
$\Gamma_{S}=\Gamma_{\nu} = 0.1\mu$ (a phenomenological damping term,
which models tachyonic preheating).
Note that the source of CP violation is here the phase between
 $\xi$ and $\zeta$, which enters at tree level.

\vspace{-0.2cm} 
\section{Contribution from Coherent Baryogenesis}
\label{contribution}
Resonant fermion production occurs whenever the fermion mode frequency 
changes nonadiabatically, which happens during the waterfall regime.
The charges stored finally in the $\chi$ fermions are displayed in the figure
below. Coupling to quarks and leptons is through the superpotential terms
\begin{equation}\label{8a}
\gamma F^c\bar{H}^cF^c\bar{H}^c/M_s \textup{~~and~~} \delta F^cH^cF^cH^c/M_s
\end{equation}
where $F^c=(\bar{4},1,2)$. We then have the decay reactions
\begin{equation}\label{9}
\chi_{1j}\rightarrow d^{c*}+\nu^{c*},\qquad\chi_{2j}\rightarrow d^c+u^c
\end{equation}
Because of the lepton number violating terms in (\ref{8a}) the charges hence get transformed to 
$q=\frac{1}{3}q_1-\frac{2}{3}q_2$. This results in a baryon to entropy ratio
\begin{equation}
  \frac{n_B}{s}  = \frac34\frac{n_B^{(0)}T_R}{V_0}
                             \approx 1\times 10^{-10}.
\label{nBs:coh.bg}
\end{equation}
where $n_B^{(0)}\simeq 1.5\times 10^{45}\,{\rm GeV}^{3}$
is the baryon density produced at preheating
(the charge density $q$ obtained from figure below  
 multiplied by $3$ colors and $1/3$, the sphaleron conversion efficiency),
$V_0 \simeq 3\times 10^{64}\,{\rm GeV}^4$
is the energy density at the end of inflation, 
$T_R = [90/(\pi^2 g_*)]^{1/4}\sqrt{\Gamma M_P}\simeq 2.7\times 10^{9}$~GeV
is the reheat temperature, $g_* = 221.5$ is the number of relativistic 
degrees of freedom of the MSSM, $M_P\simeq 2.4\times 10^{18}$~GeV 
the reduced Planck mass, $\Gamma \equiv H_R
                  \simeq 15$~GeV is the perturbative inflaton decay rate,
and $s=2\pi^2 g^*{T^3_R}/{45}$ is the entropy density.
This estimate of $T_R$ is based on the assumption that
tachyonic preheating thermalises the inflaton sector, and all other
(light) species thermalise at $T\simeq T_R$.  
The decay rate of the inflaton $\Gamma$ is dominated by the perturbative
 decay of  the $\langle \nu^c_H\rangle$-condensate into the lightest 
Majorana neutrinos with mass 
$M_1=\gamma_1\langle \nu^c_H\rangle^2_0/M_s\approx 
         4\times 10^{10}$~GeV,
such that 
$\Gamma\simeq (1/8\pi) m_{\nu_H^c}\left(\gamma_1 
 {\langle \nu^c_H\rangle_0}/{M_s}\right)^2\simeq 15$~GeV,
where $\langle\nu^c_H\rangle_0\simeq \mu$
and $m_{\nu_H^c}\simeq 4\times 10^{14}$~GeV
with our choice of parameters. 
%
\begin{figure}[htbp]
\begin{center}
\epsfig{file=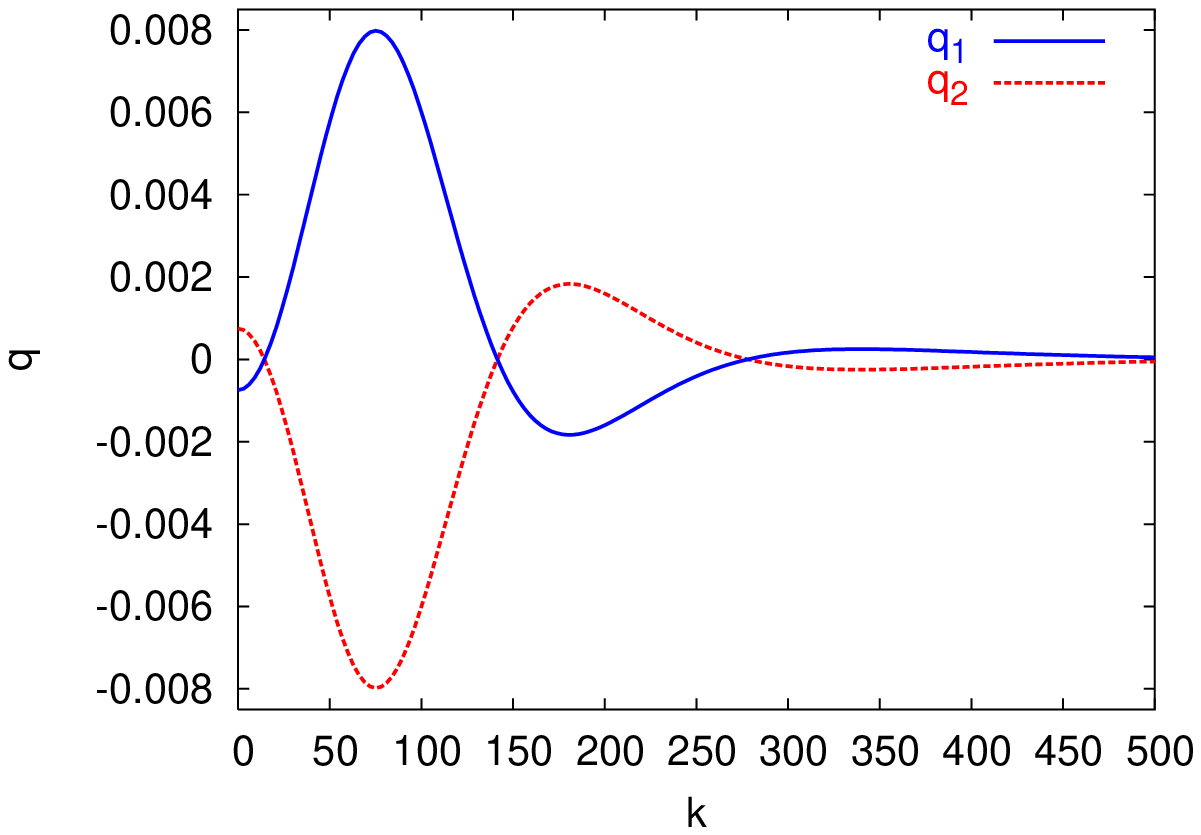, height=2.4in,width=3.5in}
\end{center}
\vskip -0.10in
\end{figure}
\noindent
The baryon to entropy ratio~(\ref{nBs:coh.bg}) is somewhat
larger than the observed value.
It can be easily reduced by chosing a smaller CP violation
(by a different choice of $\xi$ and $\zeta$), or
by reducing the reheat temperature.

\vspace{-0.2cm}
\section{Reheating and nonthermal leptogenesis}
\label{reheatingandnonthermal}
In section \ref{contribution} we have already mentioned the decay of the 
$\nu^c_H$-condensate into the lightest Majorana neutrinos after preheating. 
The reheating temperature we obtained is $T_R\simeq 2.7\times 10^9$~GeV,
which meets the gravitino bound. 
It is much below the Majorana neutrino mass $M_1\simeq 3.9\times 10^{10}$~GeV.
 The two other masses $M_{2,3}$ we assume to be much heavier.
 Thus the produced Majorana neutrino is certainly nonthermal. 
For maximal mixing and CP violation via one-loop interference like
 in thermal leptogenesis, one obtains~\cite{7}
\vspace{-0.14cm}
\begin{equation}\label{14}
\frac{n_L}{s}\leq 3\times 10^{-10}\frac{T_R}{m_{\nu^c_H}}
  \left(\frac{M_1}{10^6 {\rm GeV}}\right)
        \left(\frac{m_{\nu^3}}{0.04{\rm eV}}\right)
       \simeq 8\times 10^{-11},
\end{equation}
where we assumed the last factor to be one, 
as suggested by atmospheric neutrino oscillations. 
This gives $n_B/s \leq 3\times 10^{-11}$, which is significantly smaller than
the coherent baryogenesis result~(\ref{nBs:coh.bg}).
Increasing the cutoff scale $M_s$ lowers the value~(\ref{nBs:coh.bg}),
but reduces~(\ref{14}) at least equally fast.   


Majorana neutrinos can also be produced nonperturbatively \cite{6}, 
just like the $\chi$-particles in coherent baryogenesis. 
Nonperturbative production of the lightest Majorana neutrino
 $N_1$ is dominated by production of $N_2$ and $N_3$ with mass
 of order the inflaton mass, $M_{2,3}\simeq m_{\nu^c_H}$. 
Their decay asymmetry can be much larger than the asymmetry for $N_1$ \cite{8}.
We conclude that, like coherent baryogenesis, this production channel
 should also be regarded as a possible source for generating the BAU. 

\vspace{-0.2cm}
\section{Conclusions}
\label{conclusions}
We considered two sources for baryogenesis during (p)reheating: 
(1) Direct charge production through coherent baryogenesis;
(2) Perturbative decay of the inflaton into Majorana neutrinos. 

For the given parameters, we found coherent baryogenesis to be
 the only viable baryogenesis mechanism. For this choice of parameters the 
gravitino bound is met. For a different choice of parameters (smaller $M_s$), 
coherent baryogenesis is further enhanced, making it thus robust
baryogenesis mechanism. Since the CP-violating sources of coherent baryogenesis
arise already at tree level and not at one-loop, as for leptogenesis, 
it is a competitive source of baryons over an ample section of phase space.

%
%

\end{document}